\documentclass[12pt,a4paper]{article}

\newcommand{\rdot}{{\dot{\rho}}}
\newcommand{\sdot}{{\dot{\sigma}}}
\newcommand{\oprime}{{\omega^\prime}}
\newcommand{\tprime}{{\tau^\prime}}

\title{Bi-connected Gravity Fields\\
{\small A long due innovation}}

\author{Llu\'{\i}s\ Bel\\
\emph{wtpbedil@lg.ehu.es}
}
\begin{document}

\maketitle

\begin{abstract}

I describe a bi-connection formalism of General relativity based on the dual role of the Weitzenb\"{o}ck connection defining the parallelism at a distance and the concomitant Levi-Civita connection derived from the Riemannian metric. A more explicit tensor writing of the  geodesic and loxodromic equations clarifies their joint meaning.

\end{abstract}

\section*{Introduction}

  Given n linear independent linear forms, Cartan's geometry of linear connections includes as particular cases the so called Weitzenb\"{o}ck connection and the Levi-Civita connection. The usual formalism of General relativity uses only the Levi-Civita connection derived from a Riemannian metric. But now and then Weitzenb\"{o}ck's-like formalisms are used, \cite{Aldrovandi}, either as alternative descriptions of General relativity or to fulfiill some other purposes. In this paper I support the idea that a Weitzenb\"{o}ck like formalism should be introduced first, and the corresponding Riemannian formalism derived from it afterwards. And also that the two formalisms are not different optional descriptions of the same theory but have to be used concomitantly to make sense of the physics. I stress in particular the fact that the tensor defined by the difference of the two connections, named the Contortion, has an intrinsic meaning that strongly clarifies the meaning of the geodesy and loxodromy equations.

  I propose potentially new Field equations for vacuum that are a particular class of equations first considered by Einstein, \cite{Sauer}, pursuing other objectives. My choice has been dictated by the condition that the new formalism at the linear approximation be the same as that of General relativity at the same approximation.


\section{Generalities about Linear connections}

{\bf General Cartan connections}\,\footnote{From Chap. IV of \cite{Lichnerowicz}}
\vspace{0.5cm}

Let $V_n$ be a differential manifold of dimension $n$ and let $D_n$ and $D^\prime_n$ be two intersecting domains of $V_n$, with local coordinates $x^\alpha$ and $x^{\alpha^\prime}$, with Greek indices running from $1$ to $n$. A Linear connection on $V_n$ is a field of 1-forms of type (1,1):

\begin{equation}
\label{0.1}
\omega^\alpha_\beta(x)=\Gamma^\alpha_{\beta\gamma}(x)dx^\gamma
\end{equation}
such that at any point of the intersection of any two domains one has:

\begin{equation}
\label{0.2}
\omega^{\alpha^\prime}_{\beta^\prime}(x^\prime)=A^{\alpha^\prime}_\rho(x(x^\prime))\omega^\rho_\sigma(x(x^\prime))
A^\sigma_{\beta^\prime}(x^\prime)+A^{\alpha^\prime}_\rho(x(x^\prime))dA^\rho_{\beta^\prime}(x^\prime)
\end{equation}
with:

\begin{equation}
\label{0.3}
A^{\alpha^\prime}_\rho(x(x^\prime))=\frac{\partial x^{\alpha^\prime}}{\partial x^{\rho}}(x(x^\prime)), \quad
A^\sigma_{\beta^\prime}(x^\prime)=\frac{\partial x^\sigma}{\partial x^{\beta^\prime}}(x^\prime)
\end{equation}
where:

\begin{equation}
\label{1.3}
x^{\rho^{\prime}}=x^{\rho^{\prime}}(x),\ x^\sigma=x^\sigma(x^\prime)
\end{equation}
define the coordinate transformation.

The covariant derivatives of a function $f$ and  a covariant vector field $v_\alpha$ are by definition:

\begin{equation}
\label{0.3.1}
\nabla_\beta f=\partial_\beta f, \quad \nabla_\beta v_\alpha=\partial_\beta v_\alpha-\Gamma^\rho_{\alpha\beta}v_\rho
\end{equation}
from where, using Leibniz rule to calculate derivatives of tensor products, follows the general formula to define the covariant derivative of any tensor.

The Curvature of the linear connection $\omega^\alpha_\beta$ is the 2-form of type(1,1):

\begin{equation}
\label{0.4}
\Omega^\alpha_\beta=\frac12 R^\alpha_{\beta\gamma\delta}dx^\gamma\wedge dx^\delta
\end{equation}
where:

\begin{equation}
\label{0.5}
R^\alpha_{\beta\gamma\delta}=
\partial_\gamma\Gamma^\alpha_{\beta\delta}-\partial_\delta\Gamma^\alpha_{\beta\gamma}+
\Gamma^\alpha_{\rho\gamma}\Gamma^\rho_{\beta\delta}-\Gamma^\alpha_{\rho\delta}\Gamma^\rho_{\beta\gamma}
\end{equation}
and the Torsion is the vector-valued 2-form:

\begin{equation}
\label{0.6}
\Sigma^\alpha=\frac12 T^\alpha_{\beta\gamma}dx^\beta\wedge dx^\gamma
\end{equation}
where:

\begin{equation}\label{0.7}
T^\alpha_{\beta\gamma}=
-(\Gamma^\alpha_{\beta\gamma}-\Gamma^\alpha_{\gamma\beta})
\end{equation}

From these definitions and from the fact that $d^2=0$, $d$ being the exterior differentiation operator, there follow the following identities:

\begin{equation}
\label{0.8}
\nabla_{[\epsilon |} R^\alpha_{\beta|\gamma\delta]}=R^\alpha_{\beta\rho[\epsilon}T^\rho_{\gamma\delta]}
\end{equation}
where $[.| |..]$ is here the three indices complete anti-symmetrization operator. And:

\begin{equation}
\label{0.9}
{\nabla}_{[\alpha}{T}^\rho_{\beta\gamma]}
+{T}^\rho_{\sigma[\alpha}{T}^\sigma_{\beta\gamma]}=R^\rho_{[\alpha\beta\gamma]}
\end{equation}

The last general definition that I shall need to mention is that of an auto-parallel, which is any parameterized curve $x^\alpha(\lambda)$ solution of the differential equations:

\begin{equation}
\label{0.10}
\frac{d^2x^\alpha}{d\lambda^2}+\Gamma^\alpha_{\beta\gamma}\frac{dx^\beta}{d\lambda}\frac{dx^\gamma}{d\lambda}=a\frac{dx^\alpha}{d\lambda}
\end{equation}
where the function a in the r-h-s  depends on the parameter $\lambda$. If it is zero the parameter is said to be an affine parameter.
\vspace{0.5cm}

\vspace{1cm}
{\bf Weitzenb\"{o}ck connections}
\vspace{0.5cm}

From now on I shall use also enumeration dotted Greek indices $\rdot,\sdot,..$ running from 1 to n. $\rho$ and $\rdot$ can be contracted. For example: $\delta^\rdot_\rho=n$.

Let $\theta^\rdot_\alpha$ be $n$ linearly independent 1-forms, $e^\beta_\sdot$ being its dual contravariant vector fields:

\begin{equation}
\label{1.1}
e_\rdot^\alpha\theta_\beta^\rdot=\delta^\alpha_\beta\Leftrightarrow e^\alpha_\rdot\theta^\sdot_\alpha=\delta^\sdot_\rdot
\end{equation}

A Weitzenb\"{o}ck connection $\widetilde\Gamma^\lambda_{\beta\gamma}$ associated with  $\theta^\rdot_\alpha$ is defined as the connection that leads to the covariant derivative with symbol $\widetilde\nabla$ such that:

\begin{equation}
\label{1.1.0}
\widetilde\nabla_\alpha \theta_\beta^\rdot=\partial_\alpha \theta_\beta^\rdot-\widetilde\Gamma^\gamma_{\beta\alpha}\theta^\rdot_\gamma=0
\end{equation}
and therefore we have :

\begin{equation}
\label{1.12}
\widetilde\Gamma^\lambda_{\beta\gamma}=e^\lambda_\rdot\partial_\gamma\theta^\rdot_\beta
\end{equation}

The most significant property of Weitzenb\"{o}ck connections is that their curvature tensor is zero:

\begin{equation}
\label{1.22}
\widetilde R^\alpha_{\beta\gamma\delta}=
\partial_\gamma\widetilde\Gamma^\alpha_{\beta\delta}-\partial_\delta\widetilde\Gamma^\alpha_{\beta\gamma}+
\widetilde\Gamma^\alpha_{\rho\gamma}\widetilde\Gamma^\rho_{\beta\delta}-\widetilde\Gamma^\alpha_{\rho\delta}\widetilde\Gamma^\rho_{\beta\gamma}=0
\end{equation}

On the other hand its torsion is:

\begin{equation}
\label{Torsion}
T^\lambda_{\beta\gamma}=-(\widetilde\Gamma^\lambda_{\beta\gamma}-\widetilde\Gamma^\lambda_{\gamma\beta})=
e^\lambda_\rdot(\partial_\beta\theta^\rdot_\gamma-\partial_\gamma\theta^\rdot_\beta)
\end{equation}
And taking into account (\ref{1.22}) the identities (\ref{0.9}) become:

\begin{equation}
\label{1.25}
{\widetilde\nabla}_{[\alpha}{T}^\rho_{\beta\gamma]}
+{T}^\rho_{\sigma[\alpha}{T}^\sigma_{\beta\gamma]}=0
\end{equation}

The auto-parallels, referred to an affine parameter, are now the solutions of:

\begin{equation}
\label{Loxodromy}
\frac{d^2x^\alpha}{d\lambda^2}+\widetilde\Gamma^\alpha_{\beta\gamma}\frac{dx^\beta}{d\lambda}\frac{dx^\gamma}{d\lambda}=0
\end{equation}
and should be better called loxodromy trajectories.

\vspace{0.5cm}

{\bf Levi-Civita connections}

\vspace{0.5cm}
Let $g_{\alpha\beta}(x)$ be a Riemannian metric of any signature defined on $V_n$.

\begin{equation}
\label{metric}
ds^2=g_{\alpha\beta}(x^\lambda)dx^\alpha dx^\beta.
\end{equation}
The Levi-Civita connection with Christoffel symbols $\hat\Gamma^\rho_{\alpha\beta}$ associated with this metric define the connection which has the following two properties:

\begin{equation}
\label{Levi}
\hat\nabla_\gamma g_{\alpha\beta}=\partial_\gamma g_{\alpha\beta}-
\hat\Gamma^\rho_{\alpha\gamma}g_{\rho\beta}-\hat\Gamma^\rho_{\beta\gamma}g_{\alpha\rho}=0,
\quad \hat T^\rho_{\alpha\beta}=\hat\Gamma^\rho_{\beta\alpha}-\hat\Gamma^\rho_{\alpha\beta}=0
\end{equation}
This leads to an unique symmetric connection with symbols:

\begin{equation}
\label{1.8}
\hat\Gamma^\lambda_{\beta\gamma}=\hat\Gamma_{\beta\gamma\alpha}g^{\lambda\alpha}, \quad g^{\lambda\mu}g_{\lambda\nu}=\delta^\mu_\nu
\end{equation}
with:

\begin{equation}
\label{1.9}
\hat\Gamma_{\beta\gamma\alpha}=\frac12 (\partial_\beta g_{\gamma\alpha}
+\partial_\gamma g_{\beta\alpha}-\partial_\alpha g_{\beta\gamma})
\end{equation}

The Riemann tensor is the curvature tensor defined in (\ref{0.5}) with the corresponding general connection being substituted by the Christoffel symbols above. If it is zero then the metric $g_{\alpha\beta}$ can be reduced by a coordinate transformation to a matrix of constants $\eta_{\alpha\beta}$.

With the corresponding substitution we obtain the auto-parallels of a Levi-Civita connection:

\begin{equation}
\label{Geodesic}
\frac{d^2x^\alpha}{d\tau^2}+\hat\Gamma^\alpha_{\mu\nu}\frac{dx^\mu}{d\tau}\frac{dx^\nu}{d\tau}=0
\end{equation}
now being called the geodesics of the metric. In this case affine parameters are proportional to the proper length of the curve when this length is not zero.
\vspace{0.5cm}


\section{Connecting connections}

Let us consider any diagonal matrix whose elements $\eta_{\rdot\sdot}$ are $1$ or $-1$, the corresponding quadratic form having arbitrary signature.

To such matrix and any field of $n$ 1-forms $\theta^\rdot_\alpha$, as we considered before, it can be associated a Riemannian metric:

\begin{equation}
\label{metric}
g_{\alpha\beta}= \eta_{\rdot\sdot}\theta^\rdot_\alpha\theta^\sdot_\beta.
\end{equation}
that has the same signature as $\eta_{ab}$.

When referring to this formula I shall say that $\theta^\rdot_\alpha$ is an orthogonal  decomposition of $g_{\alpha\beta}$ and, the other way around, I shall say that $g_{\alpha\beta}$ is the metric derived from  $\theta^\rdot_\alpha$. Equivalent orthogonal decompositions:

\begin{equation}
\label{psi}
g_{\alpha\beta}= \eta_{\oprime\tprime}\theta^{\oprime}_\alpha\theta^{\tprime}_\beta.
\end{equation}
are related by point dependent frame transformations:

\begin{equation}
\label{Lorentz}
\theta^{\oprime}_\alpha=L^\oprime_\sdot(x)\theta^\sdot_\alpha(x),
\end{equation}
such that:

\begin{equation}
\label{1.17.0.2}
L^{\oprime}_{\rdot}(x)L^\tprime_\sdot(x)\eta_{\oprime\tprime}=\eta_{\rdot\sdot},
\end{equation}
modulo a permutation of the enumeration indices. Let us consider the two connections: the Weitzenb\"{o}ck connection $\widetilde\Gamma^\alpha_{\beta\gamma}$ associated with $\theta^\sdot_\alpha$ and the Levi-Civita connection $\hat\Gamma^\alpha_{\beta\gamma}$ corresponding to the Riemannian metric defined in (\ref{metric}). We have:

\begin{equation}
\label{1.17.0.3}
\hat\nabla_\gamma g_{\alpha\beta}=0, \  \widetilde\nabla_\gamma g_{\alpha\beta}=0
\end{equation}
The first is part of the definition (\ref{Levi}), and using (\ref{metric}) a short calculation proves the second. They share also a second property, namely:

\begin{equation}
\label{1.17.0.4}
\hat\nabla_\rho\eta_{\alpha_1\cdots\alpha_n}=0, \  \widetilde\nabla_\rho\eta_{\alpha_1\cdots\alpha_n}=0
\end{equation}
where $\eta_{\alpha_1\cdots\alpha_n}$ is the volume element associated with the Riemannian metric (\ref{metric})

Let us consider now the geodesic equations (\ref{Geodesic}) and the auto-parallel equations (\ref{Loxodromy}). Although these are tensor equations neither the kinetic term nor the connection dependent terms are tensors, this meaning that none of these terms have any intrinsic meaning by themselves.

Let us write the geodesic equations (\ref{Geodesic}) in the obviously equivalent form:

\begin{equation}
\label{1.21.1}
\frac{d^2x^\alpha}{d\tau^2}+\tilde\Gamma^\alpha_{\mu\nu}\frac{dx^\mu}{d\tau}\frac{dx^\nu}{d\tau}+K^\alpha_{\mu\nu}\frac{dx^\mu}{d\tau}\frac{dx^\nu}{d\tau}=0
\end{equation}
where:

\begin{equation}
\label{Contortion}
K^\alpha_{\mu\nu}=\hat\Gamma^\alpha_{\mu\nu}-\widetilde\Gamma^\alpha_{\mu\nu}
\end{equation}
is the so-called Contortion tensor of the two connections.
Define: $u^\rdot=\theta^\rdot_\alpha u^\alpha$. The geodesic equations become:

\begin{equation}\label{Force}
 \frac{du^\rdot}{d\tau}=-\theta^\rdot_\alpha K^\alpha_{\lambda\mu}u^\lambda u^\mu
\end{equation}
Noteworthy is the fact that in these equations $d /d\tau$ is an intrinsic tensor derivative  and the  right-hand term of (\ref{Force}) is an intrinsic force field.

Similarly defining $v^\rdot=\theta^\rdot_\alpha v^\alpha$ the equation of the loxodromic trajectories (\ref{Loxodromy}) of a Weitzenb\"{o}ck connection become:

\begin{equation}
\label{parallel}
\frac{dv^\rdot}{d\lambda}=0
\end{equation}
that defines  what is meant by parallel transport at a distance.

With this we succeed in giving an  intrinsic tensor meaning to the geodesic equations of the Levi-Civita connection as well as to the auto-parallels of the Weitzenb\"{o}ck connection. But it remains a fundamental indetermination due to the fact that given the n linear forms $\theta^\rdot_\alpha$ there is only one corresponding Riemannian metric while the Weitzenb\"{o}ck connection is defined only up to an event dependent orthogonal transformation (\ref{Lorentz}). I deal with this problem in the next section.


\section{Field equations (n=4)}

With 4 linear forms $\theta^\rdot_\alpha$ and $\eta_{\rdot\sdot}$ being the Minkowski metric, the formula (\ref{metric}) determines the metric $g_{\alpha\beta}$ \,\footnote{ A starting point that could lead to a generalization of General relativity briefly considered by Einstein (Sauer,(\cite{Sauer}))}. If instead the metric is known there are many compatible orthogonal decompositions. Any particular choice could be replaced by an event-dependent Lorentz transformation. But from now on I assume that only global Lorentz symmetry is accepted, this meaning that:

\begin{equation}\label{Ls}
\partial_\alpha L^\oprime_\rdot=0
\end{equation}
and therefore  everything that make sense, including (\ref{Force}) and (\ref{parallel}), is unambiguously determined.

\vspace{1cm}
{\bf The linear approximation}
\vspace{1cm}

Let  us examine carefully the problem at the linear approximation assuming that:

\begin{equation}\label{fs}
 \theta^\rdot_\alpha =\delta^\rdot_\alpha+\frac12 f^\rdot_\alpha, \quad f^\rdot_\alpha=O(1)
\end{equation}
and let us define the symmetric and anti-symmetric perturbations:

\begin{equation}\label{perturbations}
f^+_{\alpha\beta}=\frac12(\eta_{\rho\alpha}f^\rdot_\beta +\eta_{\rho\beta}f^\rdot_\alpha), \quad f^-_{\alpha\beta}=\frac12(\eta_{\rho\alpha}f^\rdot_\beta -\eta_{\rho\beta}f^\rdot_\alpha)
\end{equation}
so that:
\begin{equation}\label{fs}
 f_{\alpha\beta}=f^{+}_{\alpha\beta}+f^{-}_{\alpha\beta}
\end{equation}
The corresponding symmetric metric is then:

\begin{equation}\label{hs}
g_{\alpha\beta}= \eta_{\alpha\beta}+ f^+_{\alpha\beta},
\end{equation}
but a new field has to be considered, defined by the antisymmetric part $f^-_{\alpha\beta}$, reminiscent of Einstein's latest theory (\cite{Lichnerowicz})\,\footnote{Also: Albert Einstein,Oeuvres choisies,Seuil/CNRS, 1993)}.

A gage transformation:

\begin{equation}\label{Gage f}
 f^\rdot_\alpha\rightarrow  f^\rdot_\alpha+\partial_\alpha \zeta^\rdot
\end{equation}
induces the gage transformations:

\begin{equation}\label{Gage metric}
f^+_{\alpha\beta}\rightarrow f^+_{\alpha\beta}+ \partial_\alpha \zeta_\beta +  \partial_\beta\zeta_\alpha, \quad
f^-_{\alpha\beta}\rightarrow f^-_{\alpha\beta}+ \partial_\alpha\zeta_\beta - \partial_\beta\zeta_\alpha,
\quad \zeta_\alpha=\eta_{\alpha\rho}\zeta^\rdot
\end{equation}

 No linear combination of $\partial_\alpha f^+_{\beta\gamma}$ with the three cyclic indices exists that is gage invariant. In particular the gage transformation of the Christoffel symbols of the first kind is:

\begin{equation}\label{Cf1}
 Cf1_{\alpha\beta\,\gamma} \equiv \frac12(\partial_\alpha f^+_{\beta\gamma}+\partial_\beta f^+_{\alpha\gamma}-\partial_\gamma f^+_{\alpha\beta})\rightarrow
 Cf1_{\alpha\beta\,\gamma}+\partial_{\alpha\beta}\zeta_\gamma
\end{equation}

The first gage invariant object that we encounter is the linear approximation of the Riemann tensor:

\begin{equation}\label{Riem}
 R^{+}_{\alpha\beta\lambda\mu}=-\frac12(\partial_{\alpha\lambda}f^{+}_{\beta\mu}+\partial_{\beta\mu}f^{+}_{\alpha\lambda}
-\partial_{\alpha\mu}f^{+}_{\beta\lambda}-\partial_{\beta\lambda}f^{+}_{\alpha\mu})
\end{equation}
 or its contraction with $\eta_{\alpha\lambda}$:

 \begin{equation}\label{Rab}
 R^{+}_{\beta\mu}= \eta^{\alpha\lambda}R^{+}_{\alpha\beta\lambda\mu}
 \end{equation}

 On the other hand the object with components:

 \begin{equation}\label{Babc}
  B^{-}_{\alpha\beta\gamma}\equiv \partial_\alpha f^{-}_{\beta\gamma}+\partial_\beta f^{-}_{\gamma\alpha}+\partial_\gamma f^{-}_{\alpha\beta}
 \end{equation}
is gage invariant.

 This suggests to consider a gage invariant generalization of the linear approximation of General relativity based on these two objects (\ref{Riem}) and (\ref{Babc}). The challenge consists in choosing field equations compatible with Einstein's familiar theory as a particular case.

I proposed already a solution to this problem in a preceding paper \cite{Bel Q}. It consists in deriving the field equations from the Lagrangian:

 \begin{equation}
\label{Lagrangian}
{\cal L}=-\frac14 F^\rdot_{\alpha\beta} F^\sdot_{\lambda\mu}\eta^{\alpha\lambda}\eta^{\beta\mu}\eta_{\rho\sigma}
+\frac12 F^\rdot_{\alpha\rho}F^\sdot_{\lambda\sigma}\eta^{\alpha\lambda}+{\cal L}_{matter}
\end{equation}
where:

\begin{equation}\label{Fs}
F^{\rdot}_{\alpha\beta}=\partial_\alpha f^{\rdot}_\beta-\partial_\beta f^{\rdot}_\alpha.
\end{equation}
It leads to the  field equations:

\begin{equation}
\label{Gtensor}
G_\beta^\rdot\equiv \partial^\alpha F_{\alpha\beta}^\rdot-\delta_\beta^\rho \partial^\alpha F_\alpha +\eta^{\rdot\alpha}\partial_\alpha F_\beta=t^\rdot_\beta,
 \quad F_\alpha=F^\rdot_{\alpha\rho},\quad \partial^\beta t_\beta^\rdot=0.
\end{equation}
$t_\beta^\rdot$ being the canonical energy-momentum tensor of the field source.

Lowering the dot indices as in (\ref{perturbations}) and separating the symmetric and anti-symmetric parts, we get two sets of vacuum field equations:

\begin{equation}\label{Gs+,-}
 G^{\pm}_{\beta\gamma}= \partial^\alpha F^{\pm}_{\alpha\beta\gamma}-\eta_{\beta}\partial^\alpha \gamma F^{\pm}_\alpha +\partial_\gamma F^{\pm}_\beta=0
\end{equation}
where:

\begin{equation}\label{Fs+,-}
F^{+}_{\alpha\beta\gamma}=\partial_\alpha f^{+}_{\beta\gamma}-\partial_\beta f^{+}_{\alpha\gamma} \quad
F^{-}_{\alpha\beta\gamma}=\partial_\alpha f^{-}_{\beta\gamma}-\partial_\beta f^{-}_{\alpha\gamma}
\end{equation}

A short calculation proves that:

\begin{equation}\label{Gtensor+}
G^{+}_{\alpha\beta}= -2S^{+}_{\alpha\beta}, \quad S^{+}_{\alpha\beta}=R^{+}_{\alpha\beta}-\frac12 R^{+}\eta_{\alpha\beta}.
\end{equation}
and:

\begin{equation}
\label{Gtensor-}
G^{-}_{\beta\gamma}= \partial^{\alpha}B^{-}_{\alpha\beta\gamma}=0
\end{equation}

{\it Elementary vacuum solutions}
\vspace{1cm}

Non zero components of the linear approximation of the Schwarzschild solution:
\begin{equation}\label{Schwarzschild}
f^{+,0}_0=-\frac{m}{r},\ f^{+,1}_1=f^{+,2}_2=f^{+,3}_3=+\frac{m}{r}, \quad r=|\vec x|, \ \hbox{m constant}
\end{equation}
Non zero components of an anti-symmetric solution:

\begin{equation}\label{f(-,i}
f^{-,i}_0=\frac{m x^i}{r}, \quad i=1,2,3
\end{equation}

\vspace{1cm}
{\bf A generalization of General relativity?}
\vspace{1cm}

Let us consider the following Lagrangian:

\begin{equation}\label{Lagrangian2}
{\cal L}=-\frac14 F^\rdot_{\alpha\beta} F^\sdot_{\lambda\mu}g^{\alpha\lambda}g^{\beta\mu}\eta_{\rho\sigma}
+\frac12 F^\rdot_{\alpha\rho}F^\sdot_{\lambda\sigma} g^{\alpha\lambda}
\end{equation}
where $g_{\alpha\beta}$ is given by (\ref{metric}) and the field components are (\ref{Fs}).

Let us consider also the field equations:

\begin{equation}\label{G2}
 G^\rdot_\lambda\equiv g^{\alpha\beta}\hat\nabla_\alpha F^\rdot_{\beta\lambda}-\theta^\rdot_\lambda g^{\alpha\beta}\hat\nabla_\alpha F_\beta+\eta^{\rdot\sdot} e^\alpha_\sdot \hat\nabla_\alpha F_\lambda=0
\end{equation}
where:

\begin{equation}\label{}
e^\alpha_\sdot \theta^\sdot_\beta=\delta^\alpha_\beta, \quad F_\lambda=F^\sdot_{\lambda\mu}e^\mu_\sdot
\end{equation}
and $\hat\nabla$ is the Levi-Civita connection symbol corresponding to the metric $g_{\alpha\beta}$
At any particular event $x_0^\alpha$ we can choose a coordinate system so that:

\begin{equation}\label{x0}
(g_{\alpha\beta})_0=\eta_{\alpha\beta}, \quad (\partial_\gamma g_{\alpha\beta})_0=0
\end{equation}
which implies that at the first order of approximation at any particular event $x_0$ the Lagrangian above becomes identical with (\ref{Lagrangian}), and the Field equations above become identical with (\ref{Gtensor}).

Notice that obtaining a solution of (\ref{G2}) means to get at once the metric, the Levi-Civita and the Weitzenb\"{o}ck connections. The local Lorentz invariance of the metric has been broken by (\ref{Ls}) but the gage local symmetry of the Field variables (\ref{Fs}) has been conserved.

A second possibility is to consider the following Lagrangian:

\begin{equation}\label{Lagrangian3}
{\cal L}=-\frac14 T^\gamma_{\alpha\beta} T^\delta_{\lambda\mu}g^{\alpha\lambda}g^{\beta\mu}\eta_{\gamma\delta}
+\frac12 T^\rho_{\alpha\rho}T^\sigma_{\lambda\sigma} g^{\alpha\lambda}
\end{equation}
where now $T^\gamma_{\alpha\beta}$ is the torsion tensor (\ref{Torsion}) of the Weitzenb\"{o}ck connection.
Now the local gage symmetry

\begin{equation}\label{gage}
\theta^\rdot_\alpha\rightarrow \theta^\rdot_\alpha+\partial_\alpha \zeta^\rdot
\end{equation}
still present in (\ref{Lagrangian2}), is gone also.

In my opinion these considerations justify considering a theory based on the Lagrangian2 and Lagrangian3 above a promising generalization of General relativity. As I mentioned already, Einstein tried twice to go beyond General relativity motivated by his desire to unify gravitation and electromagnetism. The theory that I am proposing here faces a different challenge: to prove that gravitation is more than what General relativity tell us now that it is. Already, at the linear approximation, new features are puzzling: the possibility of an helicity 0 of the graviton (\cite{Bel Q}), and the existence of other solutions with $f^-_{\alpha\beta}\neq 0$.

General relativity, despite some marginal extravaganza, is a rightly glorified theory. And yet it is amazing to realize how many fine points remain to be understood.

\section*{Acknowledgements}



\begin{thebibliography}{9}

\bibitem{Lichnerowicz} A.\ Lichnerowicz, {\it Th$\acute{e}$ories relativistes de la gravitation et de l'$\acute{e}$lectromagn$\acute{e}$tisme}, Livre II, Chap. IV et V, Masson (1955)

\bibitem{Aldrovandi} R.\ Aldrovandi and J.\ G.\ Pereira, {\it Telleparallel Gravity: An introduction}, Springer, Dordrecht (2012);
www.ift.unesp.br/users/jpereira/tele.pdf
\bibitem{Sauer} T.\ Sauer, {\it Field equations in teleparallel spacetime: Einstein's Fernparallelismus approach towards unified field theory}, Historia Math. {\bf 33}, 399 (2006)

\bibitem{Pereira1} V.\ C.\ Andrade and J.\ G.\ Pereira, arXiv:gr-qc/9703059 v1 {\it Gravitational Lorentz force and the description of the gravitational interaction} (1997)
\bibitem{Pereira2} V.\ c.\ Andrade, L.\ C.\ T.\ Guillen and J.\ G.\ Pereira, arXiv:gr-qc/0011087 v2 {\it Telleparallel Gravity: An Overview} (2000)
\bibitem{Schucking2} E.\ Schucking, arXiv:gr-qc/0803.4128 v1  {\it Gravitation is torsion} (2008)
\bibitem{Bel} Ll.\ Bel, arXiv:gr-qc/0805.0846 {\it Connecting connections} (2008)
\bibitem{Schucking1} E.\ Schucking and E.\ J.\ Surowitz, arXiv:gr-qc/0703149 v1 {\it Einstein's apple: his first principle of equivalence} (2012)
\bibitem{Maluf} J.\ W.\ Maluf, arXiv:1303.3897v1 [gr-qc] {\it The teleparallel equivalent of general relativity} (2013)
\bibitem{Bel Q} Ll.\ Bel, arXiv:gr-qc/1603.01643 {\it Quantum gravity: The inverse problem} (2016)

\end{thebibliography}
\end{document}